\documentclass[prl,showpacs,preprint]{revtex4}

\usepackage{axodraw}
\usepackage{amssymb}
\usepackage{amsfonts}
\usepackage{pstricks}
\usepackage{graphicx}

\newcommand{\be}{\begin{equation}}
\newcommand{\bea}{\begin{eqnarray}}
\newcommand{\ee}{\end{equation}}
\newcommand{\eea}{\end{eqnarray}}
\def\chic#1{{\scriptscriptstyle #1}}

\begin{document}

\title{On the observability of the neutrino charge radius}

\author{J. Bernab\'eu}
\author{J. Papavassiliou}
\author{J. Vidal}

\affiliation{Departamento de F\'\i sica Te\'orica and IFIC, 
Universidad de Valencia-CSIC,\\
E-46100, Burjassot, Valencia, Spain\\}

\begin{abstract} 

It is shown that the probe-independent  
charge radius of the neutrino is a physical observable; 
as such, it may be extracted from experiment, at least in principle.  
This is accomplished by expressing a set of  
experimental $\nu_{\mu}-e$ cross-sections in terms of 
the finite charge radius and two additional  
gauge- and renormalization-group-invariant quantities,
corresponding to the electroweak effective charge 
and mixing angle. 

\end{abstract}

\pacs{12.15.Lk, 13.15.+g, 13.40.Gp, 14.60.Lm}
\preprint{FTUV-02-0603, IFIC/02-22}

\maketitle

\vspace{1.cm}

Within  the Standard  Model  the  photon ($A$) 
does  not  interact with  the
neutrino  ($\nu$) at  tree-level;   
however,  an   effective  photon-neutrino
vertex ${\Gamma}^{\mu}_{A \nu \bar{\nu}}$ 
is  generated  through  one-loop  radiative  corrections,
giving   rise   to   a   non-zero   
neutrino   charge   radius   (NCR) \cite{Bernstein:jp},
which contributes non-trivially to the 
full electron-neutrino scattering amplitude.
Even though the one-loop 
computation of the 
{\it entire} $S$-matrix element describing the aforementioned amplitude
is conceptually straighforward, the identification
of a {\it sub-amplitude},  
which would serve as the  effective 
${\Gamma}^{\mu}_{A \nu \bar{\nu}}$
has
been   faced   with  serious   complications,   associated  with   the
simultaneous   reconciliation   of   crucial  requirements   such   as
gauge-invariance,       finiteness,       and      target-independence
\cite{Lucio:1984mg}. 
The crux of the problem is that  
since in non-Abelian
gauge   theories individual   
off-shell  Green's functions are in general unphysical, the
definition of quantities familiar from scalar theories or QED, such as
effective charges and form-factors, is in general problematic.
Thus, whereas a pion form-factor may be defined perfectly well
in the one-photon approximation, 
the same definition leads to unphysical results in the case of the NCR.
The above difficulties
have    been    conclusively   settled    in
\cite{Bernabeu:2000hf},
by resorting to the well-defined electroweak gauge-invariant
separation of physical  amplitudes into effective self-energy, vertex
and  box  sub-amplitudes,   implemented  by the  pinch  technique
formalism \cite{Cornwall:1982zr}.
These  effective Green's  functions are  completely independent  of the
gauge-fixing parameter  regardless of the  gauge-fixing scheme chosen,
and  satisfy   simple,  QED-like  Ward  identities,   instead  of  the
complicated Slavnov-Taylor  identities.  The NCR
obtained  in  \cite{Bernabeu:2000hf}  :  (i)  is  independent  of  the
gauge-fixing  parameter;  (ii) is  ultraviolet  finite; (iii)  couples
electromagnetically    to   the   target;    (iv)   it    is   process
(target)-independent and  can therefore be considered  as an intrinsic
property  of  the  neutrino.   
In particular, from the gauge-invariant 
one-loop {\it proper} vertex 
$\widehat{\Gamma}^{\mu}_{A \nu_i \bar{\nu}_i}$ 
constructed using this method one extracts
the dimension-full electromagnetic
form-factor $\widehat{F}_{\nu_i}(q^2)$ as
$\widehat{\Gamma}^{\mu}_{A \nu_i \bar{\nu}_i}
= ie q^2  \widehat{F}_{\nu_i}(q^2) \gamma_{\mu}(1-\gamma_{5})$ .
The NCR, to be denoted by $\big <r^2_{\nu_i}\,  \big>$, is  
then defined as   
$\big <r^2_{\nu_i}\,  \big> = 6 \widehat{F}_{\nu_i}(0)$, and thus 
one obtains 
\be
\big <r^2_{\nu_i}\,  \big> =\, 
\frac{G_{\chic F}}{4\, {\sqrt 2 }\, \pi^2} 
\Bigg[3 
- 2\log \Bigg(\frac{m_{\chic i}^2}{M_{\chic W}^2} \Bigg) \Bigg]\, ,
\,\,\,\,\,\,\,\, i= e,\mu,\tau
\label{ncr}
\ee
where $m_i$ denotes the mass of the charged iso-doublet
partner of the neutrino under consideration, and $G_{\chic F}$
is the Fermi constant. 
The numerical values of the NCR given in  Eq.(\ref{ncr}) 
are  
$\big< r^2_{\nu_e}\, \big> =
4.1 \times 10^{- 33} $ cm${}^2$, 
$\big< r^2_{\nu_{\mu}}\, \big> = 
2.4 \times 10^{- 33}$ cm${}^2$, and 
$\big< r^2_{\nu_{\tau}}\, \big> =
1.5 \times 10^{- 33}$ cm${}^2$.  
The   
classical definition of the NCR 
(in the static limit) as the second moment 
of the  spatial neutrino charge density $\rho_{\nu}(\bf{r})$, i.e. 
$\big <r^2_{\nu}\,  \big> = e^{-1} \int d {\bf{r}} r^2 \rho_{\nu}({\bf{r}})$, 
suggests the heuristic interpretation of the above numbers 
as a measure of the ``size'' of the neutrino $\nu_i$ when probed 
electromagnetically.

The unambiguous resolution of the theoretical 
issues which was accomplished in \cite{Bernabeu:2000hf}, together with 
the definite numerical predictions quoted above, 
inevitably leads to the next important
questions: Can the NCR be measured, even in principle ?
Does it qualify as a ``physical observable'' ?
In this Letter we will show that the answer to the
above questions is affirmative. 

It is important to clarify from the outset what we mean by
``measuring'' the NCR, especially in light of the
fact that bounds on the NCR already appear in the literature
\cite{Salati:1994tf}. 
From our point of view, 
measuring the entire process  
$f^{\pm}\nu \to f^{\pm} \nu$ does 
{\it not} constitute a measurement
of the NCR, because by 
 changing the target fermions $f^{\pm}$ one will 
generally change the answer, thus introducing a target-dependence
into a quantity which (supossedly) constitutes an
intrinsic property of the neutrino. 
Instead, what we want to measure is the target-independent
Standard Model NCR only, 
stripped of any target dependent contributions. 
Specifically, as mentioned above, the PT
rearrangement of the $S$-matrix
makes possible the definition of distinct, physically meaningful
sub-amplitudes, one of which,   
$\widehat{\Gamma}^{\mu}_{A \nu_i \bar{\nu_i}}$, 
is finite and directly
related to the NCR. 
However, the remaining sub-amplitudes, such as 
self-energy, vertex- and box-corrections,  
even though they 
do no enter into the definition of the
NCR, still contribute numerically to the entire $S$-matrix; 
in fact, some of them combine to form additional  
physical observables of interest, 
most notably the effective (running) 
electroweak charge and mixing angle.
Thus, in order to isolate the NCR, one must conceive of
a combination of experiments and kinematical conditions, 
such that 
all contributions not related to the NCR
will be eliminated. 

In this paper we propose a set of such 
(thought) experiments
involving neutrinos
and    anti-neutrinos.         
Consider the elastic processes 
$ f \nu  \to f \nu $ and 
$f \bar{\nu}  \to f \bar{\nu} $,
where $f$ denotes an electrically charged 
fermion belonging to a different
iso-doublet than the neutrino $\nu$, in order to eliminate 
the diagrams mediated by a charged $W$-boson.
The Mandelstam variables are defined as
$s=(k_1+p_1)^2 = (k_2+p_2)^2$, 
$t= q^2 = (p_1-p_2)^2 = (k_1-k_2)^2$, 
$u = (k_1-p_2)^2 = (k_2-p_1)^2$, 
and $s+t+u=0$ (see Fig.1). 
In what follows we will restrict ourselves to the 
limit $t=q^2 \to 0$ of the above amplitudes,
assuming that all external (on-shell) fermions are massless.
As a result of this special kinematic situation we have the
following relations:
$p_1^2 = p_2^2 = k_1^2 = k_2^2 = p_1 \cdot p_2 = k_1 \cdot k_2 = 0$
and 
$p_1 \cdot k_1  = p_1 \cdot k_2 = p_2 \cdot k_1 = p_2 \cdot k_2 = s/2 $.
In the center-of-mass system we have that 
$t=-2 E_{\nu}E_{\nu}'(1-x)\leq 0 $, 
where $E_{\nu}$ and $E_{\nu}'$
are the energies of the neutrino before and after the
scattering, respectively, and 
$x \equiv \cos\theta_{cm}$, where
$\theta_{cm}$
is the scattering angle. Clearly, the condition $t=0$ 
corresponds to the exactly forward amplitude, 
with $\theta_{cm}=0$, \, $x=1$. 
Equivalently, in the laboratory frame,
where the (massive) target fermions are at rest, the 
condition of $t=0$ corresponds to the kinematically extreme 
case where the target 
fermion remains at rest after the scattering. 

\begin{figure}[!t]
\includegraphics[width=10.5cm]{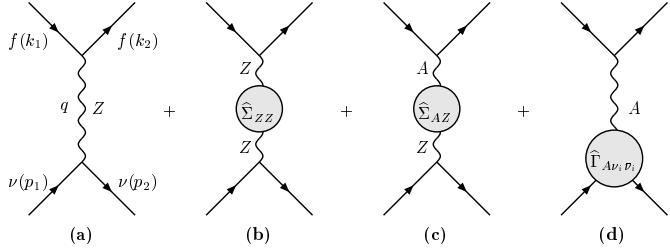}
\caption{\label{figure} The universal (a-c) and flavour-dependent (d)
contributions to $\sigma^{(+)}_{\nu f}$.}
\end{figure} 

At tree-level the amplitude  $ f \nu  \to f \nu $ is 
mediated by an off-shell $Z$-boson, coupled to the fermions  
by means of the bare vertex 
$\Gamma_{Z {f} \bar{f}}^{\mu} = -i 
(g_w/ c_w)\, \gamma^{\mu}\, [ v_f + a_f \gamma_5]$
with 
$v_f = s^2_w Q_{f} - \frac{1}{2} T^f_z$  and 
$a_f=\frac{1}{2} T^f_z$;
$Q_f$ is the electric charge of the fermion $f$, 
$T^f_z$ its $z$-component of the weak iso-spin,  
$c_w = \sqrt{1 - s^2_w} = M_{\chic W}/M_{\chic Z}$, 
the electric charge $e$ is related to the $SU(2)_L$ 
gauge coupling $g_w$ by $e=g_w s_w$.
At one-loop, the relevant  
contributions may be unambiguously determined  
through the standard pinch technique 
rearrangement of the amplitude, giving rise to 
gauge-independent sub-amplitudes. 
In particular, the one-loop 
$AZ$ self-energy $\widehat{\Sigma}_{\chic{A}\chic{Z}}^{\mu\nu}(q^2)$
obtained is transverse, for {\it both} 
the fermionic and the bosonic contributions,
i.e. it may be written in terms of the 
dimension-less scalar function 
${\widehat{\Pi}}_{ \chic{A} {\chic Z}} (q^2)$ as
$\widehat{\Sigma}_{\chic{A}\chic{Z}}^{\mu\nu}(q^2)
= (q^2 \, g^{\mu\nu}  - q^{\mu} q^{\nu}) 
{\widehat{\Pi}}_{ \chic{A} {\chic Z}} (q^2)$.
Of course, the $ZZ$ self-energy 
$\widehat{\Sigma}_{\chic{Z}\chic{Z}}^{\mu\nu}(q^2)$
is not transverse; in what follows we will 
discard all longitudinal pieces, since they vanish between 
the conserved currents of the massless external fermions,
and will 
keep only the part proportional to $g^{\mu\nu}$, 
whose dimension-full cofactor will be denoted 
by $\widehat{\Sigma}_{\chic{Z}\chic{Z}}(q^2)$.
If the fermion mass $m$ were non-vanishing, the 
longitudinal pieces would  
induce additional terms proportional 
to positive powers of $(m/M_W)$ and/or  $(m/\sqrt{s})$;
the former are naturally suppressed, whereas the 
latter may be made arbitrarily small, by 
adjusting appropriately the value of $s$.  
Furthermore, as is well-known, 
the one-loop
vertex 
$\widehat\Gamma_{{\chic Z} {\chic F} \bar{\chic F}}^{\mu}(q,p_1,p_2)$, 
with $F = f$  or $F = \nu$, 
satisfies a QED-like 
Ward identity, relating it to the one-loop  
inverse fermion propagators $\widehat\Sigma_{\chic F}$,
i.e  
$ q_{\mu} 
\widehat\Gamma_{{\chic Z} {\chic F} \bar{\chic F}}^{\mu}(q,p_1,p_2)
= 
\widehat\Sigma_{\chic F} (p_1) - \widehat\Sigma_{\chic F} (p_2)$.
It is then easy to show that, in the limit of 
$q^2 \to 0$,  
$\widehat\Gamma_{{\chic Z} {\chic F} \bar{\chic F}}^{\mu} 
\sim q^2 \gamma^{\mu}(c_1 + c_2 \gamma_5)$; 
since it is multiplied by a
massive $Z$ boson propagator $(q^2 - M_{\chic Z})^{-1}$, its 
contribution to the amplitude vanishes when 
$q^2 \to 0$. This is to be contrasted with the 
$\widehat{\Gamma}^{\mu}_{A \nu_i \bar{\nu}_i}$,  
which is accompanied by a 
$(1/q^2)$ photon-propagator, thus giving rise 
to a contact interaction between the target-fermion and the neutrino,
described by the NCR. 
 
We next proceed to eliminate the target-dependent box-contributions;
to accomplish this we resort to the ``neutrino--anti-neutrino'' method.
The basic observation 
is that the tree-level amplitudes 
${\cal M}_{\nu f}^{(0)}$ 
as well as the
part of the one-loop amplitude ${\cal M}_{\nu f}^{(B)}$
consisting of 
the propagator and vertex corrections 
(the ``Born-improved'' amplitude),
are proportional to $[\bar{u}_{f}(k_2)\gamma_{\mu}( v_f + a_f \gamma_5 ) 
u_{f}(k_1)]
[\bar{v}(p_1)\gamma_{\mu} P_{\chic L} \, v(p_2)]$, 
and  
therefore transform differently than the boxes 
under the replacement 
$\nu \to \bar{\nu}$. 
In particular, 
the coupling of the 
$Z$ boson to a pair of on-shell anti-neutrinos 
may be written in terms of on-shell neutrinos
provided that
one changes the chirality projector from 
$P_{\chic L} = \frac{1}{2} (1-\gamma_5)$ to 
$P_{\chic R} = \frac{1}{2} (1+\gamma_5)$, 
and supplying a relative minus sign \cite{Sarantakos:1983bp}, i.e.
\bea
\bar{v}(p_1) \, \Gamma_{Z\bar{\nu}\bar{\nu}}\,  v(p_2) &=& 
i \bigg(\frac{g_w}{2c_w}\bigg)\bar{v}(p_1)\gamma_{\mu} 
P_{\chic L} \, v(p_2)\nonumber\\ 
&=& - \, 
i \bigg(\frac{g_w}{2c_w}\bigg)\bar{u}(p_2)\gamma_{\mu} P_{\chic R} \, u(p_1)
\label{Gantinu}
\eea
To obtain the above results, 
we simply use the fact that since the quantities considered
are scalars in the spinor space 
their values coincide with those of their transposed, and 
employ subsequently 
$
\gamma_{\mu}^{T} = - C \gamma_{\mu}  C^{-1}$, 
$\gamma_{5}^{T} = C \gamma_{5}  C^{-1}$, 
$v^{T}(p) C  = \bar{u}(p)$, 
$C^{-1} \bar{v}^{T}(p) = u(p)$,
where $C$ is the charge conjugation operator.
Thus, under the above transformation, 
${\cal M}_{\nu f}^{(0)} + {\cal M}_{\nu f}^{(B)}$ reverse 
sign once, 
whereas the box contributions reverse sign twice.
These distinct transformation properties 
allow for the isolation of
the box contributions 
when judicious combinations of the forward differential cross-sections
$(d\sigma_{\nu f}/dx)_{x=1}$ and  
$(d\sigma_{\bar{\nu} f}/dx)_{x=1}$ 
are formed. In particular, 
$\sigma^{(+)}_{\nu f} \equiv 
(d\sigma_{\nu f}/dx)_{x=1}
+ (d\sigma_{\bar{\nu} f}/dx)_{x=1}$ does not contain 
boxes, i.e. 

\be
\sigma^{(+)}_{\nu f} 
= \frac{1}{16 \pi s}  
\Bigg[{\cal M}_{\nu f}^{(0)} *
{\cal M}_{\nu f}^{(0)\dagger} + 
2\, \Re e \bigg({\cal M}_{\nu f}^{(0)}* 
{\cal M}_{\nu f}^{(B)\dagger }
\bigg)\Bigg]
\label{Plus}
\ee
whereas the conjugate combination 
$\sigma^{(-)}_{\nu f} \equiv 
(d\sigma_{\nu f}/dx)_{x=1}
- (d\sigma_{\bar{\nu} f}/dx)_{x=1}$  
isolates the contribution of the boxes.
The $*$ in the above formulas 
denotes that the trace over initial and final fermions
must be taken.

Finally, a detailed analysis \cite{JJJ} shows 
that in the kinematic limit we consider, 
the Bremsstrahlung contribution vanishes, 
due to a  a completely destructive interference 
between the two relevant  diagrams corresponding to the
processes $f A \nu (\bar{\nu}) \to f \nu (\bar{\nu})$ and 
$f \nu (\bar{\nu}) \to f A \nu (\bar{\nu})$. 
The absence of such corrections is consistent with the
fact that there are no infrared divergent contributions 
from the (vanishing) vertex
$\widehat\Gamma_{{\chic Z} {\chic F} \bar{\chic F}}^{\mu}$, 
to be cancelled against.  

From Eq.(\ref{Plus}) and Fig.1 we see that 
$\sigma^{(+)}_{\nu f}$ receives contributions from the 
tree-level exchange of a $Z$-boson (Fig.1a), the one-loop contributions
from the ultraviolet divergent quantities 
$\widehat{\Sigma}_{\chic{Z}\chic{Z} }(0)$ and  
${\widehat{\Pi}}^{ \chic{A}  {\chic  Z}} (0)$ (Fig.1b and Fig.1c, 
respectively),  
and the (finite) NCR, coming from the proper vertex 
$\widehat{\Gamma}^{\mu}_{A \nu_i \bar{\nu}_i}$, (Fig.1d).
The first three contributions are universal, i.e. common to all 
neutrino species, whereas that of the proper vertex 
$\widehat{\Gamma}^{\mu}_{A \nu_i \bar{\nu}_i}$
is flavor-dependent. As a consequence, 
the flavor-dependent part of the NCR can 
be immediately separated out by taking in $\sigma^{(+)}_{\nu f}$ 
the difference for two neutrino species. 
In particular, for the case of $\nu_{\mu}$ and  $\nu_{\tau}$,
we obtain from Eq.(\ref{Plus})
\be
\sigma^{(+)}_{\nu_{\mu}\, e } - \sigma^{(+)}_{\nu_{\tau}\, e } 
= \lambda\, (1- 4 s_w^2)\,
\bigg(\big< r^2_{\nu_{\mu}}\, \big> - 
\big< r^2_{\nu_{\tau}}\, \big>\bigg)
\label{RIJ}
\ee
where $\lambda \equiv (2\sqrt{2}/3) s \alpha \,G_{\chic F}$, 
$\alpha = e^2/4\pi$ is the fine-structure constant.
A priori, the difference in the 
forward amplitudes ${\cal M}_{\nu_{\mu} e} - {\cal M}_{\nu_{\tau} e}$  
would contribute to a difference for the neutrino index of refraction 
\cite{Botella:1986wy} in electron matter; this difference vanishes, 
however, for ordinary matter due to its neutrality.

Next we will demonstrate that one can actually do better than that,  
obtaining from experiment not only the difference but even 
the absolute value of the NCR for a given neutrino flavour.
To discuss this methodology, the renormalization of 
$\widehat{\Sigma}_{\chic{Z}\chic{Z} }(0)$ and  
${\widehat{\Pi}}_{ \chic{A}  {\chic  Z}} (0)$ must be carried out.
It turns out that, by virtue of the Abelian-like Ward-identities 
enforced after the pinch technique rearrangement 
\cite{Cornwall:1982zr},
 the resulting expressions combine in such a way as to form manifestly 
renormalization-group invariant combinations 
\cite{Hagiwara:1994pw,Papavassiliou:1996fn}.
In particular, after 
carrying out the standard re-diagonalization \cite{Baulieu:ux},
two such quantities 
may be constructed (see third paper in 
\cite{Papavassiliou:1996fn}):
\bea
\bar{R}_{\chic{Z}}(q^2) &=& \frac{1}{4\pi} \, \bigg(\frac{g_w}{c_w}\bigg)^2
\bigg[q^2 - M_\chic{Z}^2 +\Re e\,\{\widehat{\Sigma}_{\chic{Z}\chic{Z}}(q^2)\}
\bigg]^{-1}
\nonumber\\  
\bar{s}_w^{2}(q^2) &=&  s_w^{2}\Biggl(1 - \frac{c_w}{s_w}\, 
\Re e\,\{\widehat{\Pi}_{\chic{A}\chic{Z}}(q^2)\}\Biggr) \,.
\label{RW}
\eea
where $ \Re e\,\{...\}$ denotes the real part. 
These quantities retain the same form  
when written in terms 
of unrenormalized or renormalized quantities,  
due to the special conditions enforced on the renormalization 
constants, analogous to the text-book QED relation $Z_1=Z_2$
between the renormalization constants of the 
vertex and the fermion self-energy. 
In addition to being renormalization-group invariant, 
both quantities defined in  Eq.(\ref{RW})
are universal (process-independent); 
$\bar{R}_{\chic{Z}}(q^2)$ corresponds to the $Z$-boson 
effective charge, while $\bar{s}_w^{2}(q^2)$  
corresponds to an effective mixing angle. 
We emphasize that 
the  renormalized  ${\widehat{\Pi}}_{ \chic{A}  {\chic  Z}} (0)$
{\it cannot}  form part of the NCR,  because 
it fails  to form a
renormalization-group invariant
quantity on its own.  
Thus, if 
${\widehat{\Pi}}_{ \chic{A}  {\chic  Z}} (0)$ were to be considered 
as the ``universal'' 
part of the NCR, to be added to   
the finite and flavour-dependent contribution 
comming  from the  proper vertex,  then  
the  resulting NCR would 
depend on  the subtraction point and scheme chosen
to renormalize it,
and would therefore be unphysical.    
Instead,
${\widehat{\Pi}}_{\chic{A} {\chic Z}}  (0)$ must be combined with the
appropriate tree-level 
contribution (which  evidently  does not  enter into  the
definition of the NCR, since it is $Z$-mediated) 
in order to form the effective  
$\bar{s}_w^{2}(q^2)$ 
acting on the electron vertex,
whereas the finite 
NCR will be determined from the proper neutrino vertex only. 

After recasting $\sigma^{(+)}_{\nu f}$ of 
 Eq.(\ref{Plus})
in terms of manifestly 
renormalization-group invariant building blocks, one may 
fix $\nu = \nu_{\mu}$, 
and then consider three different choices for $f$: (i) 
right-handed electrons, $e_{\chic R}$; 
(ii) left-handed electrons, $e_{\chic L}$, and (iii) neutrinos, 
$\nu_{i}$ 
other than  $\nu_{\mu}$, i.e. $i=e,\tau$. 
It is then straighforward to verify from Eq.(\ref{Plus}) and  
Eq.(\ref{RW}) that 
$\bar{R}^2(0)$ is directly written in terms of the physical 
cross-section $\sigma^{(+)}_{\nu_{\mu} \,\nu_i}$, as
\be
\sigma^{(+)}_{\nu_{\mu} \,\nu_i} = s \pi \bar{R}^2(0)
\ee 
This cross-section constitutes 
a fundamental ingredient for neutrino propagation
in a neutrino medium \cite{Notzold:1987ik}, and is relevant for 
astrophysical and cosmological scenarios. 
Similarly, for the electron target we obtain the system
\bea
\sigma^{(+)}_{\nu_{\mu} \,e_{\chic R}}  &=& 
s \pi \bar{R}^2(0)\, \bar{s}_w^{4}(0) 
- 2 \lambda s_w^{2} \, 
\big< r^2_{\nu_{\mu}}\, \big>
\nonumber\\
\sigma^{(+)}_{\nu_{\mu} \,e_{\chic L}} &=& 
s \pi \bar{R}^2(0) \,
\bigg(\frac{1}{2} - \bar{s}_w^{2}(0)\bigg)^{2}   
+ \lambda (1-2 s_w^{2}) \, 
\big< r^2_{\nu_{\mu}}\, \big> 
\label{syst1}
\eea
At this point one possibility would be to 
extract {\it indirectly} the value of the NCR, 
using the 
precision electroweak predictions  
for $\bar{R}^2(0)$ and  
$\bar{s}_w^{2}(0)$  \cite{Hagiwara:1994pw}
as input in 
Eq.(\ref{syst1}). Much better, there is a second possibility, 
whereby $\bar{R}^2(0)$, $\bar{s}_w^{2}(0)$, and  
$\big< r^2_{\nu_{\mu}}\, \big>$ are treated as three unknown 
quantities, to be determined from the above equations. 
This procedure, although more involved,  allows 
(at least conceptually) for a {\it direct} 
measurement of NCR. 
Substituting  $s \pi \bar{R}^2(0) \to \sigma^{(+)}_{\nu_{\mu} \,\nu_i}$
into  Eq.(\ref{syst1}), we arrive at 
a system which is linear in the unknown quantity 
$\big< r^2_{\nu_{\mu}}\, \big>$, and 
quadratic in $\bar{s}_w^{2}(0)$. 
The corresponding solutions are given by
\bea
\bar{s}_{w}^{2}(0) &=&  s_w^{2} \pm \Omega^{1/2}
\nonumber\\
\big< r^2_{\nu_{\mu}}\, \big> &=&  
\lambda^{-1}
\Bigg[\bigg(s_w^{2}-\frac{1}{4} \pm \Omega^{1/2}\bigg)
\sigma^{(+)}_{\nu_{\mu} \,\nu_i}  
+ \sigma^{(+)}_{\nu_{\mu} \,e_{\chic L}} - 
\sigma^{(+)}_{\nu_{\mu} \,e_{\chic R}}
\Bigg]\,
\label{sol2}
\eea
where the discriminant $\Omega$ is given by
\be
\Omega = (1- 2 s_w^{2}) \bigg(\frac{\sigma^{(+)}_{\nu_{\mu} \,e_{\chic R}}}
{\sigma^{(+)}_{\nu_{\mu} \,\nu_i}} -  
\frac{1}{2} s_w^{2}  
\bigg)
+  2 s_w^{2} 
\frac{\sigma^{(+)}_{\nu_{\mu} \,e_{\chic L}}}
{\sigma^{(+)}_{\nu_{\mu} \,\nu_i}}
\label{disc}
\ee
and must satisfy $\Omega > 0$. 
The actual sign in front of $\Omega$ may be chosen by requiring that 
it correctly accounts for the sign of the shift of $\bar{s}_{w}^{2}(0)$
with respect to $ s_w^{2}$ predicted by the theory \cite{Hagiwara:1994pw}. 

To extract the experimental values of the quantities 
$\bar{R}^2(0)$, $\bar{s}_w^{2}(0)$, and $\big< r^2_{\nu_{\mu}}\, \big>$,
one must substitute  
in Eq.(\ref{sol2}) and Eq.(\ref{disc}) the experimentally
measured values for the differential cross-sections 
$\sigma^{(+)}_{\nu_{\mu} \,e_{\chic R}}$, 
$\sigma^{(+)}_{\nu_{\mu} \,e_{\chic L}}$,
and $\sigma^{(+)}_{\nu_{\mu} \,\nu_i}$. 
This means that to solve the system one would 
have to carry out three different pairs of experiments. 

The theoretical values of the $\bar{R}^2(0)$ and $\bar{s}_w^{2}(0)$
are obtained from  Eq.(\ref{RW}).
Since (by construction) these two 
quantities are renormalization-group invariant, one may choose  
any renormalization scheme for computing their value.
In the ``on-shell'' (OS) scheme \cite{Sirlin:1981yz}
the experimental values for the 
input parameters $s_w$ and $\alpha$ are 
$s_w^{\chic{(OS)}}  = 0.231$ 
and $\alpha^{\chic{(OS)}} = 1/128.7$;  
the renormalized self-energies 
$\widehat{\Sigma}_{\chic{Z}\chic{Z}}^{\chic R} (q^2)$ and 
$\widehat{\Pi}_{\chic{A}\chic{Z}}^{\chic R}(q^2)$ are defined as 
$\widehat{\Sigma}_{\chic{Z}\chic{Z}}^{\chic R}(q^2) = 
\widehat{\Sigma}_{\chic{Z}\chic{Z}}(q^2) - 
\widehat{\Sigma}_{\chic{Z}\chic{Z}}(M^2_Z) - 
\, (q^2 - M^2_Z) \widehat{\Sigma}'_{\chic{Z}\chic{Z}}(q^2)|_{q^2=M^2_Z}$,
where the prime denotes differentiation with respect to $q^2$, 
and   
$\widehat{\Pi}_{\chic{A}\chic{Z}}^{\chic R}(q^2) =
\widehat{\Pi}_{\chic{A}\chic{Z}}(q^2) - 
\widehat{\Pi}_{\chic{A}\chic{Z}}(M^2_Z)$. Substituting 
in the resulting expressions (see for example \cite{Hagiwara:1994pw})
standard values for the quark and lepton masses, and choosing for 
the Higgs boson a mass $M_H = 150$ GeV, we obtain 
$\bar{R}^2(0) = 1.86\times 10^{-3}/M^4_Z $ 
and $\bar{s}_w^{2}(0) = 0.239$.

To summarize, we have found that the interaction of 
$\nu_{\mu}$'s with other neutrino species and with 
left- and right-handed
electrons provides at $q^2=0$ a definite framework for separating out 
the probe-independent NCR from other gauge- and 
renormalization-group-invariant quantities, i.e. the effective electroweak 
charges $\bar{R}^2(0)$ and $\bar{s}_w^{2}(0)$. 
 The analysis has used the 
symmetric combination of neutrinos and anti-neutrinos to avoid 
contributions from box-diagrams. Once the observable character 
of the NCR has been established, we plan to extend the method to the 
entire electromagnetic form-factor analysis by means of the 
coherent neutrino-nuclear scattering \cite{JB}. 
Finally note that, for the Dirac neutrinos that we consider, 
the neutrino anapole moment \cite{Lucio:1984mg} 
is simply equal to $\frac{1}{6} \big< r^2_{\nu}\, \big>$, 
due to the $(1-\gamma_5)$ character of the vertex. 
Therefore, all 
theoretical properties of the NCR, as well as its 
observability, carry over automatically to this quantity as well. 

{\it Acknowledgments}:
This paper has been supported by the Grant AEN-99/0692 
of the Spanish CICYT. \\

\end{document}